\documentclass[preprint,preprintnumbers,amsmath,amssymb]{revtex4}

\usepackage{graphicx}
\usepackage{dcolumn}
\usepackage{bm}

\begin{document}


\title{\bf Predicting morphotropic phase boundary locations and transition temperatures in Pb- and Bi-based perovskite solid solutions from crystal chemical data and first-principles calculations}

\author{Ilya Grinberg}

\address{The Makineni Theoretical Laboratories, Department of Chemistry, University of Pennsylvania, Philadelphia, PA 19104-6323}
\author{  Matthew R. Suchomel and  Peter K. Davies }
\address{ Department of Materials Science and Engineering, University of Pennsylvania, Philadelphia, PA 19104-6323}
\author{ and Andrew M. Rappe}
\address{ The Makineni Theoretical Laboratories, Department of Chemistry, University of Pennsylvania, Philadelphia, PA 19104-6323}

\date{\today}

\begin{abstract}
Using data obtained from first-principles calculations, we show that the position of the morphotropic phase boundary (MPB) and transition temperature at MPB in ferroelectric perovskite solutions can be predicted with quantitative accuracy from the properties of the constituent cations.   We find that the mole fraction of PbTiO$_3$ at MPB in Pb(B$'$B$''$)O$_3$-PbTiO$_3$, BiBO$_3$-PbTiO$_3$ and Bi(B$'$B$''$)O$_3$-PbTiO$_3$ exhibits a linear dependence on the ionic size (tolerance factor) and the ionic displacements of the B-cations as found by density functional theory calculations.   This dependence is due to competition between the local repulsion and A-cation displacement alignment interactions.  Inclusion of first-principles displacement data also allows accurate prediction of transiton temperatures at the  MPB. The obtained structure-property correlations are used to predict morphotropic phase boundaries  and transition temperatures in as yet unsynthesized solid solutions.

\end{abstract}

\maketitle

\section{\label{sec:level1}Introduction }
Perovskite solid solutions have been used extensively in ferroelectric and piezoelectric applications.  For the past few decades, the PbZr$_{1-x}$Ti$_x$O$_3$  (PZT) solid solution has been the material of choice. The highest PZT piezoelectric  response is found for compositions in the vicinity of the morphotropic phase boundary (MPB).  In the search for better fundamental understanding of these complex systems and for better technological properties, many other Pb-based solid solutions have been investigated~\cite{Ye02p35,Noheda02p27,Park97p1804,Eitel01p5999,Randall04p3633,Suchomel04p4405,Landolt}.  Currently a large amount of data is available in the literature on the phase diagrams and piezoelectric properties of a wide variety of binary and ternary Pb-based solid solutions.
  Nevertheless, materials with better electromechanical properties and/or higher transition temperature have been obtained only recently due to advances in single crystal growth methods~\cite{Park97p1804} and due to the alloying of Bi ions on the perovskite A-site~\cite{Eitel01p5999,Randall04p3633,Suchomel04p4405}.   The search for better materials has been  difficult, due to the lack of guidance for predicting MPB locations  and transition temperatures.  
This necessitates time consuming experimental synthesis for  a range of compositions to locate the MPB and to determine the properties of interest.   While first-principles calculations are of great use in elucidating the structure and interatomic interactions in known materials~\cite{Cohen92p136,Grinberg02p909,Bellaiche00p5427,Singh95p12559,Fornari01p092101,Cockayne99pR12542,Saghi-Szabo99p12771} and DFT calculations are now faster than experimental synthesis, their utility in predicting new ferroelectric materials is weakened by an inexact representation of the real material by an approximate exchange-correlation functional.  This often leads to  errors in lattice constant of one to two  percent~\cite{Wu04p104112}.  Since ferroelectric materials are very sensitive to pressure, even such relatively small errors in volume can lead to dramatic inaccuracies in computed properties.   Thus, the search for new piezoelectrics is currently guided by empirical rules based on the competition between the preferred A-O and B-O bond lengths.  This is encapsulated in the tolerance factor $t$

\begin{eqnarray}
 t = \frac {R_{\rm A-O}}  {R_{\rm B-O} \sqrt 2},
\end{eqnarray}

\noindent where $R_{\rm A-O}$ is the sum of A and O ionic radii  and $R_{\rm B-O}$ is the sum of B and O ionic radii.  For example, the tolerance factor has been widely used in solid state chemistry to predict stability of the perovskite structure.  (A tolerance factor lower than 0.9 or higher than 1.1 typically makes the perovskite structure unstable due to a mismatch  between preferred A-O and B-O bond lengths.)  Recently, in a major advance, Eitel {\em et al.} suggested an empirical relationship~\cite{Eitel01p5999} between tolerance factor and the transition temperature at the MPB which has guided the discovery of high $T_c$ alloy BiScO$_3$-PbTiO$_3$ (BS-PT).  This empirical rule states that as the tolerance factor of the non-PT end member decreases, the transition temperature at the MPB ($T_c^{\rm MPB}$) increases.   

The discovery of high $T_c^{\rm MPB}$ in BSPT has motivated an intense experimental effort in synthesis and characterization of new Bi-based piezoelectrics~\cite{Duan04p2185,Inaguma04p231,Stringer05p024101,Suchomel04p4405}.  It was found that while the $t$-$T_c^{\rm MPB}$ empirical rule captures the general trend,  comparison of $T_c^{\rm MPB}$ of Bi(Mg$_{1/2}$Ti$_{1/2}$)O$_3$-PT (BMT-PT) and Bi(Mg$_{1/2}$Zr$_{1/2}$)O$_3$-PT (BMZ-PT) revealed significant deviation.   Whereas the  $t$-$T_c^{\rm MPB}$ relationship of Ref. 4 predicts that $T_c^{\rm MPB}$  of BMZ-PT would be 80~K higher than $T_c^{\rm MPB}$ of BMT-PT, experimental results show that $T_c^{\rm MPB}$ of BMZ-PT  is actually lower by 125~K than $T_c^{\rm MPB}$ of BMT-PT~\cite{Suchomel04p4405}.

Ref. 6 considers a qualitative correlation between the tolerance factor of the non-PT end member and the mole fraction of PT at the MPB ($x_{\rm PT}^{\rm MPB}$).
They showed that while there is a large scatter in $x_{\rm PT}^{\rm MPB}$ for higher tolerance factor values (between 1.02  and 0.95), the relationship becomes more quantitative  for  non-PT end members with lower $t$ values (0.95 to 0.88).

Although the work of Refs. 4 and 6 provide some  important guidelines for the search for new materials, several questions remain.  First, what is the underlying physical reason for these empirical rules?  Second, can better predictors of $x_{\rm PT}^{\rm MPB}$ and $T_c^{\rm MPB}$ be devised based on the understanding of the mechanisms underlying  compositional phase transitions and trends in Curie temperatures?
In this work, we show that tolerance factor is a good predictor of the $x_{\rm PT}^{\rm MPB}$  due to the relationship between the local structure driven A-B cation interactions and phase transitions of ferroelectric perovskite solid solutions.  We also  show that by taking into account the off-centering tendencies of the B-cations as revealed by DFT calculations, a more accurate predictive relationship between properties of the constituent ions and $x_{\rm PT}^{\rm MPB}$ and $T_c^{\rm MPB}$ can be obtained to guide future synthesis efforts.

\section{\label{sec:level1} Methodology}
{\em Ab-initio} density functional theory (DFT) calculations were performed for the (1-$x$)Pb(Sc$_{1/2}$Nb$_{1/2})$O$_3$--$x$PT (PSN-PT) and (1-$x$)Pb(In$_{1/2}$Nb$_{1/2})$O$_3$--$x$PT (PIN-PT) 
systems using $2 \times 2 \times 2$ 40-atom  supercells with periodic boundary
conditions at experimental volume~\cite{Landolt}.   The energy of the system was evaluated using a local density
approximation exchange-correlation functional~\cite{Perdew81p5048} and
was minimized with respect to the atomic coordinates by a quasi-Newton
method with no symmetry imposed.  A $2
\times 2\times 2$ $k$-point sampling of the Brillouin zone was used.  The
calculations were done with designed non-local optimized norm conserving pseudopotentials~\cite{Ramer99p12471,Rappe90p1227}.
We study five compositions with $x = 0$, 0.25, 0.5, 0.75  and 1.0.
  For pure PSN  and PIN we used a 40-atom supercell with  B-cations in a rock-salt arrangement.  
Three B-cation arrangements with minimal oxygen over- and under-bonding~\cite{Cockayne99pR12542,Grinberg04p144118, Grinberg04p220101} were used to study $x$=0.25, $x$=0.5 and $x$=0.75 compositions. 
We find that the values for cation displacement vary little among the supercells studied.   The data for PZT, Pb(Mg$_{1/3}$Nb$_{2/3}$)O$_3$-PT (PMN-PT), Pb(Zn$_{1/3}$Nb$_{2/3}$)O$_3$-PT  (PZN-PT) and Pb(Sc$_{2/3}$W$_{1/3}$)O$_3$-PT systems were taken from previous studies~\cite{Grinberg04p144118,Juhas04p214101,Grinberg04p220101}.
The excellent agreement between the pair distribution functions (PDF) obtained from our relaxed DFT structures~\cite{Grinberg04p144118,Juhas04p214101,Grinberg04p220101}  and by neutron-scattering experiments~\cite{Juhas04p214101,Egami98p1,Egami01p33} for PMN, PZT and PSW  indicate  that our calculations accurately portray the local structure of these materials.

\section{\label{sec:level1} Results}

\subsection{\label{sec:level2} Morphotropic Phase Boundary Locations}

The correlation between $x_{\rm PT}^{\rm MPB}$ and the tolerance factor of the end-member directly reflects the  relationship between local interatomic interactions and compositional phase transitions. In PT, the tetragonal phase is strongly favored, resulting in a high $c$/$a$ ratio.  The tetragonality is driven by  off-center distortions on the A-site~\cite{Halilov02p3443,Halilov04p174107}.  Perovskites where only the B-site cations are ferroelectrically active (e.g., BaTiO$_3$ and KNbO$_3$) have rhombohedral ground states.  DFT studies have shown that the cation distortions in PbTiO$_3$  are coupled;  in the absence of Ti distortion, large Pb  off-centering is unfavorable, and conversely large Ti distortions are unfavorable if the Pb atom is held in the center of its oxygen cage~\cite{Ghosez99p836}.  This indicates that a strong Pb-Ti repulsive interaction is present. The existence of such an interaction is not surprising, since both Pb and Ti are positively charged.  Since there is no bonding between Pb and Ti, an overlap of Pb and Ti wavefunctions  gives rise to Pauli repulsion only.   Besides the direct Pb-Ti repulsion, bond-order conservation at the oxygen atoms tends to  make short Pb-O and Ti-O bonds to the same oxygen atom unfavorable; this leads  to an indirect  oxygen-mediated Pb-Ti repulsion~\cite{Grinberg04p144118}.  Similar coupling between A-site lone pair and B-site distortions were also noted in a wide range of complex oxides~\cite{Halasyamani04p3586}.

 For Pb-based perovskites with more than one B-cation, DFT calculations show that Pb atoms will try to avoid the larger B-cation and move toward the smaller one~\cite{Grinberg02p909,Grinberg04p144118}.   This is due to a larger wavefunction overlap and a stronger repulsion with the larger ion.  
This behavior is illustrated in Figure 1 for the 50/50 PZT solid solution, with Pb ions moving away from the Zr-rich faces and toward the Ti-rich faces.  The stronger Pb-Zr repulsion results in different positions of the first peak in the Pb-Zr and Pb-Ti partial PDFs (Figure 2a).   Similarly, comparison of Pb-In and Pb-Sc partial PDFs obtained from DFT calculations of PSN and PIN shows that first Pb-In peak is located at larger distances than the first Pb-Sc peak (Figure 2b).  The differences in peak positions for Zr and Ti (0.11~\AA) and In and Sc (0.05~\AA) are similar to the Shannon-Prewitt~\cite{Shannon76p751} ionic radius differences between Zr and Ti (0.12~\AA) and In and Sc (0.055~\AA).  In a solid solution, the (100) cube faces that Pb atoms move toward can be classified as either high, medium or low Pb-B repulsion depending on the size of the B-cations located at the cube face corners.  For example, in the case of the PZT solid solutions, DFT calculations show that 3 or 4 Zr atoms in a (100) face disfavors Pb displacements, leading to either a smaller magnitude of Pb distortion (thus increasing Pb-Zr distance and lessening the Pb-Zr repulsion) or a tilt in the distortion direction toward a more friendly cube face with smaller number of Zr atoms.    At the MPB,  the Pb-B-cation repulsion cost induced by collinear (100) Pb distortions is equal to the cost of disordering the Pb distortions (due to dipole-dipole and bonding interactions). At higher Zr content, Pb cations distort in a variety of low symmetry directions toward available low Zr-content faces.   This results in a disordered structure with overall (111) polarization~\cite{Grinberg02p909,Grinberg04p144118,Corker98p6251,Egami01p33}.   

This mechanism implies that a solid solution of PT and a perovskite with a larger size B-cation such as  Pb(Yb$_{1/2}$Nb$_{1/2}$)O$_3$ (PYN) (average B-cation size of 0.754~\AA\ versus 0.72~\AA\ for Zr) will have an MPB at higher PT content due to stronger Pb-B-cation repulsion and a large energy cost of collinear (100) displacements.  Conversely, a solid solution of PT and a perovskite with a smaller size B-cations such as PMN (average B-cation size of 0.67~\AA\ versus 0.72~\AA\ for Zr) will have an MPB at lower PT content due to weaker Pb-B-cation repulsion.\footnote[1]{In the following arguments, we ignore the influence of indirect through-oxygen Pb-B-cation repulsions.  While these interactions are important in ordered systems such as PbMg$_{1/3}$Nb$_{2/3}$O$_3$ and PbZn$_{1/3}$Nb$_{2/3}$O$_3$ (see Ref. 23), we expect that that under-bonding and overbonding induced by the presence of heterovalent cations on the B-site will average out for compositions relevant to this work, due to destruction of B-cation order by Ti substitution at low PT content.}
. Similarities between Bi and Pb behavior on the perovskite A-site mean that this mechanism will not be fundamentally changed by the presence of Bi ions on the A-site.  We quantify these arguments by 

\begin{eqnarray}
 \Delta E_{T-R}(x_{\rm nPT}) = E_{T}(x_{\rm nPT}) -  E_{R} (x_{\rm nPT}) 
\end{eqnarray}

\noindent and 

\begin{eqnarray}
 \Delta E_{T-R}(x_{\rm nPT}) = -\Delta E_{\rm dis}(x_{\rm nPT}) + \Delta E_{\rm A-B} (x_{\rm nPT}) 
\end{eqnarray}

\noindent where $x_{\rm nPT}$ is the fraction of the non-PT end member in the solid solution, $\Delta E_{T-R}(x_{\rm nPT})$ is the energy difference between the $T$ and $R$ phases,  $\Delta E_{\rm dis}(x_{\rm nPT})$ is the energy cost of disordering the cation displacements due to electrostatic and bonding interaction and $\Delta E_{\rm A-B} (x_{\rm nPT})$ is the energy cost due to A-B repulsion incurred by collinear (100) cation displacements of Pb and Bi ions.    The MPB is located at $x_{\rm nPT}^{\rm MPB}$ when $\Delta E_{T-R}(x_{\rm nPT})$ is 0 and $\Delta E_{\rm dis}(x_{\rm nPT}^{\rm MPB})$ = $\Delta E_{\rm A-B} (x_{\rm nPT}^{\rm MPB})$.  
Since Pb  displacements in the tetragonal phase  of PT solid solutions are fairly similar (Table I), we can make a zeroth-order approximation that $\Delta E_{\rm dis}(x_{\rm nPT})$  is independent of composition and is constant for all systems.  Although this approximation is very rough and ignores the much larger magnitude of Bi displacements, it turns out to be valid for reproducing experimental observations.

  As a first approximation, we take the A-B repulsion to be  a linear function of the ionic size of the B-cation and the A-B interatomic distance.   The dependence of the $\Delta E_{\rm A-B} (x)$ on ionic size is intuitive -- larger ions give rise to stronger interatomic repulsions. A large ionic size of the B-cation usually makes B-cation displacements unfavorable (rattling cation model)~\cite{Egami00p16,Slater50p748}, giving rise to larger indirect repulsion contributions as well.     Since Pb displacements in the tetragonal phase and the perovskite volume are fairly constant for a wide variety of PT solid solutions, the Pb-B interatomic distance will be a linear function of the B-cation displacement.     

Analysis of Bi-B-cation distances shows that the dependence  of Bi-B-cation repulsion on ionic size and displacement of the B-cation will not be too different from the case of Pb-B-cation repulsive interactions.
For an ideal perovskite structure with a cube edge $a$ $\approx$ 4.0~\AA, the A-B-cation distance is  3.46~\AA. While a 0.45~\AA\ Pb distortion creates a Pb-B-cation distance of 3.23~\AA\ (assuming no B-cation displacement), the much larger 0.8~\AA\ Bi distortion~\cite{Iniguez03p224107} creates a Bi-B-cation distance  of 3.07~\AA.  The difference between these two distances is  0.16~\AA\, which is similar to the difference between the Pb$^{2+}$ and Bi$^{3+}$ ionic radii (1.49~\AA\ for Pb$^{2+}$ and 1.36~\AA\ for Bi$^{3+}$).  In other words, the smaller ionic size of Bi allows it to move much closer to the B-cations before a further Bi displacement becomes unfavorable due to a rapidly rising Bi-B-cation repulsion.  Since Pb and Bi are similar chemically, we expect that the Bi-B-cation repulsion will exhibit a similar dependence on interatomic distance and B-cation ionic size.  The strength of the A-B repulsion is thus affected in the same manner by the B-cation size and off-centering for  both Pb and Bi occupation of the A-site; this  makes a fairly accurate treatment of both cases with one equation possible.  We write

\begin{eqnarray}
 \Delta E_{\rm A-B} (x_{\rm nPT}) = x_{\rm nPT} (a_0+b_0 R_{\rm B}^{\rm avg}-c_0 D_{\rm B}^{\rm avg} ) +(1-x_{\rm nPT}) (a_0+b_0 R_{\rm Ti}-c_0 D_{\rm Ti})
\end{eqnarray}

\noindent where $a_0$, $b_0$ and $c_0$ are constants, $R_{\rm B}^{\rm avg}$ is the average ionic size of the B-cations in the non-PT end member as computed from the B-cation composition and Shannon-Prewitt ionic radii, and $D_{\rm Ti}$ and  $D_{\rm B}^{\rm avg}$  are distortion magnitudes for the Ti and B-cations of the non-PT end member respectively.  The assumption that  $\Delta E_{T-R}$ is a linear function of composition is supported by the experimental data for the PZT solid solution~\cite{Rossetti02p8131} (Figure 3).
  Examination of the  data presented in Table I shows that displacements for a particular B-cation in the tetragonal phase are similar among different solid solutions (Figure 4). This means that $D_{\rm B}$ is a transferable property of the B-O$_6$ complex. Using the displacement magnitudes obtained by DFT calculations, we therefore assign $D_{\rm B}$ values for Mg, In, Sc, W, Zr, Fe, Nb, Zn, and Ti cations (Figure 9).

At the MPB, $R$ and $T$ phase energies are equal and

\begin{eqnarray}
\Delta E_{T-R} (x)= 0 = x_{\rm nPT} (a_0+b_0 R_{\rm B}^{\rm avg}-c_0 D_{\rm B}^{\rm avg} ) +(1-x_{\rm nPT}) (a_0+b_0 R_{\rm Ti}-c_0 D_{\rm Ti}) -\Delta E_{\rm dis} 
\end{eqnarray}

\noindent Solving for $x_{\rm nPT}$, setting the numerator to 1 and collecting all terms that do not depend on the identity of non-PT end member, we get, using new constants $a_1$,$b_1$ and $c_1$ 

\begin{eqnarray}
x_{\rm nPT} =  1/(a_1  +b_1 R_{\rm B}^{\rm avg} - c_1 D_{\rm B}^{\rm avg}).
\end{eqnarray}

\noindent  Expanding this expression in a Taylor series to first order, we get

\begin{eqnarray}
x_{\rm nPT} =  a_2 - b_2 R_{\rm B}^{\rm avg} +c_2 D_{\rm B}^{\rm avg},
\end{eqnarray}

\noindent where $a_2$, $b_2$ and $c_2$ are new constants resulting from the Taylor expansion.  Since the PT content at MPB ($x^{\rm MPB}_{\rm PT}$) is equal to 1-$x^{\rm MPB}_{\rm nPT}$, we get

\begin{eqnarray}
x^{\rm MPB}_{\rm PT} =  1- a_2  + b_2 R_{\rm B}^{\rm avg}  - c_2 D_{\rm B}^{\rm avg}
\end{eqnarray}

\noindent  
Since a large average size of end member B-cations is equivalent to a lower end member tolerance factor, we can write

\begin{eqnarray}
x^{\rm MPB}_{\rm PT} =  a_3  - b_3 t_{\rm nPT}  - c_3 D_{\rm B}^{\rm avg},
\end{eqnarray}

\noindent  where we once again use a Taylor series to convert from $R_{\rm B}^{\rm avg}$ to $t_{\rm nPT}$ and collect all terms that do not vary with $t_{\rm nPT}$ or $R_{\rm B}^{\rm avg}$ into the constant $a_3$.  

The dependence of $x^{\rm MPB}_{\rm PT}$ on $t_{\rm nPT}$ in Eqn. 9 explains the empirical relationship between $t$ and $x_{\rm PT}^{\rm MPB}$ found in Ref. 6.  Examination of the plot of $x_{\rm PT}^{\rm MPB}$ vs $t_{\rm nPT}$   (Table II, Fig 5) shows that a narrow band of about 0.2 width encompasses most points, with outliers lying below the lower edge. (PbMn$_{2/3}$W$_{1/3}$O$_3$ is the exception to this trend, but perhaps the presence of the much larger Mn$^{2+}$ ions makes the tolerance factor assigned to this system incorrect as discussed below).     For some cases the deviations are quite large, e.g. for PZN and PbFe$_{1/2}$Nb$_{1/2}$O$_3$ (PFN).  For solid solutions with $x_{\rm PT}^{\rm MPB}$ values lying in the narrow band, no more than two thirds of the end member B-cations are occupied by ferroelectrically active Nb and Fe ions and no strongly off-centering Zn and Ti cations are present.
For these solid solutions, the  $c_3  D_{\rm B}^{\rm avg}$ term is fairly constant, and the position of the MPB shows a good correlation with the  tolerance factor of the non-PT end member.  On the other hand, examination of the outlier compositions reveals that the B-site is either fully occupied by Fe$^{3+}$ and Nb$^{5+}$ B-cations or contains Zn and Ti cations.  The large Nb, Zn and Ti displacements found in our DFT calculations  and the 0.17~\AA\ Fe$^{3+}$ displacement found by DFT calculations~\cite{Wang03p1719} on BiFeO$_3$, mean that $D_{\rm B}^{\rm avg}$  values for these solutions  are significantly higher than  the typical $D_{\rm B}^{\rm avg}$ values found for the data points  in the narrow band.    Thus, the deviations from the  $t$-$x_{\rm PT}^{\rm MPB}$ relationship are due to differences in the average B-cation displacements.  Sorting the solid solutions by the content of ferroelectrically active cations on the  non-PbTiO$_3$ end-member B-site, we see that the data now falls into two distinct groups with the lower $x^{\rm MPB}_{\rm PT}$ values for the end-members with high content of ferroelectrically active cations (Figure 5).

We now quantify the above qualitative analysis.  Since correlations based on Eqn. 8-9 will hold only as long as the changes in the ionic displacements and radii are not too large and the Taylor expansion converges, we choose the functional form in Eqn. 5 in order to treat the full range of possible PT solid solutions.  Using  $R_{\rm B}$ and $D_{\rm B}$ values (Table III) for solid solutions containing elements for which we have DFT  $D_{\rm B}$ data (Table I),  we fit the experimentally observed PT content at MPB to equation 6 to obtain a new predictor of $x^{\rm MPB}_{\rm PT}$ :

\begin{eqnarray}
x^{\rm MPB}_{\rm PT} =  1 -1/(-0.32+4.53 R_{\rm B}^{\rm avg}  -7.99 D_{\rm B}^{\rm avg}).
\end{eqnarray}

\noindent   The correlation between the $x^{\rm MPB}_{\rm PT}$ values predicted by the fit  and  observed experimentally is shown in Fig 6.  All predicted PT mole fractions are within 0.15 for the end members with B-cations for which we have DFT displacement data.

Examination  of the correlation in Figure 6 shows that Eqn. 10 reproduces general trends in  MPB locations for solid solutions with only Pb on the A-site as well as for Bi(B$'$B$''$)O$_3$-PT solid solutions with both Pb and Bi ions on the A-site.  This confirms that while the smaller ionic size of Bi decreases Bi-B-cation repulsion making the tetragonal phase more stable, the larger Bi displacement has the opposite effect of almost the same magnitude.

As a check for obtained results, we evaluate $x^{\rm MPB}_{\rm PT}$ for PT end member itself.  Clearly, mixing PT with PT will never result in an MPB.  Therefore, a prediction of a positive value of  $x^{\rm MPB}_{\rm PT}$ by Eqn. 10  would indicate a failure of our model. In fact, using $R_{\rm Ti}$ and $D_{\rm Ti}$ values from Table II, we get $x^{\rm MPB}_{\rm PT}$=-1.47.  This means that PT is tetragonal, as observed experimentally.

Despite the lack of  DFT displacement data, simple arguments can be used to estimate reasonable $D_{\rm B}$ values for    Co$^{2+}$,  Ni$^{2+}$, Cd$^{2+}$, Mn$^{3+}$ and Yb$^{3+}$ cations.  These can then be used to further test the validity and the predictive usefulness of Eqn 10.  
 The ionic size  of Ni$^{2+}$, Co$^{2+}$, Cd$^{2+}$ and Yb$^{3+}$ (in the high-spin state typically found in perovskites) is similar to or larger than that of Mg$^{2+}$ cation.  According to the rattling cation model, such a large size will make distortions unfavorable as the large B-cations do not have room to off-center.  Unlike the Zn$^{2+}$ cation, Ni$^{2+}$, Co$^{2+}$ and Yb$^{3+}$ do not have empty low lying $p$  electronic states that enable covalent bonding and distortion of Zn$^{2+}$ in the Zn-O$_6$ cage~\cite{Grinberg04p220101,Halilov02p3443}.  In the case of Cd$^{2+}$ (where such low lying states are present), the extremely large ionic radius (0.95~\AA) will prevent significant off-centering.  We therefore estimate $D_{\rm B}$ values for these cation to be equal to 0.08~\AA\ obtained for Mg$^{2+}$ by DFT calculations.  The smaller Mn$^{3+}$  has more room to off-center and should be more ferroelectrically active  with $D_{\rm B}$ value  higher than that for Mg$^{2+}$.   The lower Ti content at MPB for PFN-PT than for PbMn$_{1/2}$Nb$_{1/2}$O$_3$-PT (PMnN-PT) despite the identical size of the Fe$^{3+}$ and Mn$^{3+}$ cations, implies that $D_{\rm Mn}$ should smaller than $D_{\rm Fe}$ (0.17 \AA).  Based on experimental MPB data for Fe and Mn containing systems, we estimate $D_{\rm Mn}$ of 0.12 \AA.


The $x_{\rm PT}^{\rm MPB}$ values generated for Pb(B$'$B$''$)O$_3$-PT solid solutions based on the assigned displacements (Table III, Figure 6) are generally in good agreement with experimental values~\cite{Landolt,Suchomel04p4405}.  However, large disagreement is found for Pb(Mn$_{2/3}$W$_{1/3}$)-PT solution as well as a general underestimation of $x_{\rm PT}^{\rm MPB}$ for BiBO$_3$-PT and BiB$'$B$''$O$_3$-PT systems and a general overestimation of  $x_{\rm PT}^{\rm MPB}$ for Pb-based perovskites.  In the case of Pb(Mn$_{2/3}$W$_{1/3}$)-PT it is possible that this discrepancy (which is also present in Figure 5) is due to the presence of Mn$^{2+}$ ion in the system.   While the difference  between Pb(Mn$_{2/3}$W$_{1/3}$)O$_3$ and Pb(Mn$_{1/2}$W$_{1/2}$)O$_3$ Mn mole fractions is fairly small (0.16), the impact of the change of Mn$^{3+}$ into Mn$^{2+}$ is significant because of the large difference in ionic radii of the two Mn oxidation states (0.83~\AA\ for Mn$^{2+}$ and 0.645~\AA\ for Mn$^{3+}$).  The larger ionic radius of Mn$^{2+}$ gives rise to smaller tolerance factor and to a higher PT content at the MPB than that predicted based on Mn$^{3+}$ ionic size.  In fact, 0.55 MPB PT mole fraction  predicted for the PbMn$_{1/2}$W$_{1/2}$O$_3$-PT solution by Eqn. 10 is quite close to the experimental value of 0.53.  

The underestimation of $x_{\rm PT}^{\rm MPB}$ for BiBO$_3$-PT and BiB$'$B$''$O$_3$-PT systems suggests that  Bi substitution on the A-site gives rise to a change in dependence of the $\Delta E_{\rm A-B}$ on the average ionic size and displacement of the B-cation.   Although this change is not large as indicated by the moderate size of the  deviations of predicted $x_{\rm PT}^{\rm MPB}$ from experimental values, the effect of large Bi displacement is not fully canceled out by a smaller Bi ionic size, leading to stronger A-B  repulsive interactions and underestimation of $x_{\rm PT}^{\rm MPB}$ and overestimation of  $x_{\rm PT}^{\rm MPB}$ for Pb-based perovskites.   

To take the difference between Pb and Bi into account, we now fit Eqn. 6 separately for Pb-based solid solutions and for solutions with mixed Pb/Bi occupation on the A-site and obtain

\begin{eqnarray}
x^{\rm MPB}_{\rm PT} =  1 -1/(0.34+3.31 R_{\rm B}^{\rm avg}  -7.49 D_{\rm B}^{\rm avg}).
\end{eqnarray}

\noindent for solid solutions with only Pb on the A-site, and 

\begin{eqnarray}
x^{\rm MPB}_{\rm PT} =  1 -1/(-0.97+5.71 R_{\rm B}^{\rm avg}  -7.69 D_{\rm B}^{\rm avg}).
\end{eqnarray}

\noindent for solid solutions with Pb and Bi on  the A-site.
    This improves the correlation between the predicted and experimental $x_{\rm PT}^{\rm MPB}$ values (Figure 7). We therefore expect that  Eqns. 11-12 will be better predictors of $x_{\rm PT}^{\rm MPB}$ values than Eqn. 10.

\subsection{\label{sec:level2} Transition Temperatures at the MPB}

We find that incorporation of the first-principles displacement data leads to more accurate predictions of $T_c$  at the MPB.   Eitel {\em et al.} have  shown that despite some scatter in the experimental data, a lower tolerance factor of the non-PT end-member generally leads to a higher $T_c^{\rm MPB}$.   Joint experimental and DFT studies of Pb-based solid solutions have shown that ferroelectric transition temperatures in perovskite systems are proportional to the square of Pb and B-cation displacements~\cite{Juhas04p214101,Grinberg04p220101} and are enhanced by coupling between nearest neighbor ferroelectrically active cations~\cite{Juhas04p2086}.   We therefore investigate if the scatter in the $t$-$T_c^{\rm MPB}$ correlation is due to the differences in B-cation displacements.   

Examination of the experimental $T_c^{\rm MPB}$ data (Table IV, Figure 8) shows that the data fall into two groups.  Both groups exhibit a linear dependence of $T_c^{\rm MPB}$ on $t$ but the  $T_c^{\rm MPB}$  values are shifted down for BS-PT, BMZ-PT and BiMg$_{3/4}$W$_{1/4}$O$_3$-PT (BMW-PT), PSW-PT, PbCo$_{1/2}$W$_{1/2}$O$_3$-PT (PCoW-PT) and PbMg$_{1/2}$W$_{1/2}$O$_3$-PT (PMW-PT), relative to $T_c^{\rm MPB}$  values  for the other  solid solutions. 
  This split is due to the differences in B-cation compositions between the two groups.  The end-member perovskites in the first group (from now on referred to as group 1) do not contain any strongly ferroelectric Nb, Fe, Ti or  Zn ions on the B-site.  Instead, the B-site is fully occupied by weakly ferroelectric cations Mg, Co, Sc and W in BS, BMW, PCoW and PSW end-members.  In BMZ end-member, one half of the B-site is occupied by weakly ferroelectric Mg cation and one half of the B-site is occupied by the Zr cation, which displays strong off-centering only in solid solutions with high PT concentration.   On the other hand, for the end-member perovskites in the second group (from now on referred to as group 2) either at least half of the B-site is occupied by the strongly off-centering Nb, Fe, Ti or Zn ions or the absence of strongly off-centering ions in the non-PT end-member is balanced by the absence of the weakly ferroelectric ions in the case of  PbZrO$_3$-PT, PbHfO$_3$-PT and PbSnO$_3$-PT solutions

The dependence of $T_c^{\rm MPB}$ values on  ferroelectric activity of the B-cations is consistent with the correlation between  Curie temperatures and the square of ionic displacements in ferroelectric perovskites~\cite{Abrahams68p551,Juhas04p214101,Grinberg04p220101}.  Similarly, a correlation between a high $T_c$ and low $t$ (which has been noted previously~\cite{Eitel01p5999}) can be explained as follows.  Small $t$ values indicate that the B-site cation prefers  a larger volume than the A-site cations.  Such an expansion of the A-site has been shown to lead to increased A-site displacements for the ferroelectrically active Pb and Bi A-site cations~\cite{Halilov02p3443,Tinte03p144105}.  Since transition temperatures have been shown to correlate with the square of ionic displacements, a small tolerance factor will lead to a higher transition temperature.    
Especially small end-member tolerance factors are achieved for Bi-based perovskites, due to the small size of Bi$^{3+}$ cation.  
The larger separation of Bi lone pairs as compared to Pb lone-pairs~\cite{Hyde89book,Seshadri01p487} may also contribute to the high $T_c$ values observed for Bi-based ferroelectrics.

Carrying out linear regression analysis of the $T_c^{\rm MPB}$ data for the two groups of solid solutions\footnote[3]{In the regression analysis for group 1, we omit data for BMW-PT solid solution due to the large experimental  error bar in $T_c^{\rm MPB}$ (See Ref. 18).}  we obtain 

\begin{eqnarray}
T_c^{\rm MPB} = 4511 - 4514 t
\end{eqnarray}

\noindent for solid solutions in group 1, and

\begin{eqnarray}
T_c^{\rm MPB} =  6634 - 6539 t  
\end{eqnarray}

\noindent for solid solutions in group 2,  where $T_c^{\rm MPB}$ is in $^{\circ}$C. The fit to data in group 1 is quite similar to the extrapolation used by  by Eitel {\em et al.}  to predict $T_c^{\rm MPB}$ in BS-PT, BiInO$_3$-PT (BI-PT) and BiYbO$_3$-PT (BY-PT) solid solutions.  This is consistent with the weak ferroelectric behavior of Sc$^{3+}$, In$^{3+}$ and Yb${^3+}$ ions.  The relatively low scatter of the data in Fig. 8 means that  Equations 13 and 14 can be used to accurately screen for perovskite solid solutions with high  $T_c^{\rm MPB}$  values.

\section{\label{sec:level1} Discussion}

The strong correlation exhibited in Figure 7  suggests that the dependence of the MPB position on the ionic size and displacements should hold for as yet unexamined Pb and Bi-based solid solutions. Below we discuss some of the  solutions which should exhibit interesting properties at the MPB (Table V). Among Pb-based binary perovskites, the PbZn$_{1/2}$W$_{1/2}$O$_3$ (PZW)  and PbCd$_{1/2}$W$_{1/2}$O$_3$ (PCdW) double perovskites and their solid solution with PbTiO$_3$ are the most promising systems for experimental investigation.   Unlike the  PbB$^{3+}_{1/2}$B$^{5+}_{1/2}$O$_3$ perovskites, all  PbB$^{2+}_{1/2}$B$^{6+}_{1/2}$O$_3$ perovskites exhibit a rocksalt B-cation arrangement due to a high value of disordering energy created by large charge difference of +2 and +6 cations.   These compounds are typically  antiferroelectric; the ferroelectric rhombohedral phase only appears after substantial (more than 0.2) Ti substitution and is followed by a tetragonal phase~\cite{Landolt}.  For PZW, substitution of the highly ferroelectrically active Zn$^{2+}$ on the B$^{2+}$ site should have an impact similar to Ti substitution.  At the very least, this will shift the AFE-FE transition as well as the  MPB to lower $x_{\rm PT}^{\rm MPB}$ value (Table V).  It is also possible that the end member itself will be ferroelectric in the ground state due to total B-cation ordering and  the partial occupation of the B-site by ferroelectrically active cations.    For the PCdW-PT solid solution, Eqn.13  predicts that  $T_c^{\rm MPB}$ (487 $^{\circ}$C) will be higher than that of PZT (385 $^{\circ}$C).

For BiB$'$B$''$O$_3$-PT solid solutions, Eqn. 12 predicts  that  a combination of Zn on the B$'$ site and Ti, Nb or W cations on the B$''$ site will require very low PT substitution to stabilize the tetragonal phase.   Using $R_{\rm B}^{\rm avg}$  and $D_{\rm B}^{\rm avg}$ values in Table II, we get $x^{\rm MPB}_{\rm PT}$ values of -0.07, 0.26 and 0.21 for BiZn$_{1/2}$Ti$_{1/2}$O$_3$, BiZn$_{2/3}$Nb$_{1/3}$O$_3$ and BiZn$_{3/4}$W$_{1/4}$O$_3$ respectively.  The negative value of $x^{\rm MPB}_{\rm PT}$ for BiZn$_{1/2}$Ti$_{1/2}$O$_3$ implies that mixing  BiZn$_{1/2}$Ti$_{1/2}$O$_3$ with PbTiO$_3$ will never result in an MPB, with tetragonal phase being preferred for all compositions. This prediction is consistent with our recently obtained experimental data for BZT-PT solid solutions~\cite{Suchomel05preprint}  (Figure 9).

One of the main goals of current research in piezoelectrics is the discovery of solid solutions with high $T_c$  at the MPB.  For the new systems to be technologically useful, it must be possible to synthesize a pure perovskite phase of the material under ambient conditions.     
 While extremely low tolerance factor end-members (e.g. BiYbO$_3$) would give rise to high PT content at MPB and a high $T_c^{\rm MPB}$, the instability of the perovskite phase in the pure end member  make the synthesis of such  solid solutions challenging~\cite{Eitel01p5999,Duan04p2185,Inaguma04p231}.  
BiB$'$B$''$O$_3$ end-members with high ferroelectric cation content provide an alternative route for achieving high MPB transition temperatures without sacrificing perovskite stability.  For BiCd$_{1/2}$Ti$_{1/2}$O$_3$ (BCdT), BiZn$_{1/2}$Zr$_{1/2}$O$_3$ (BZZ), BiZn$_{2/3}$Nb$_{1/3}$O$_3$ (BZN)  and BiZn$_{3/4}$W$_{1/4}$O$_3$ (BZW) end-members,  Eqn. 14 predicts $T_c^{\rm MPB}$ values of $\approx$ 770~$^{\circ}$C, 640~$^{\circ}$C,  580~$^{\circ}$C and 570~$^{\circ}$C, respectively.  These compare favorably with $T_c^{\rm MPB}$ values of 507~$^{\circ}$C and 615~$^{\circ}$C for BI and BY end-members respectively, predicted by Eqn. 13. 
 For BZW and BZN solid solutions, Eqn. 11  predicts a rather low PT content at MPB.   It has been shown experimentally that for BZN-PT solid solution,  pure perovskite phase stability extends only to 30$\%$ BZN content\cite{Nomura82p1471}. This will make the synthesis of the MPB composition difficult by conventional solid-state chemistry methods. 
  For  BCdT end-member, $x^{\rm MPB}_{\rm PT}$ value is 0.52.  Since the BS-PT solid solution is stable in the perovskite phase up to 0.5 BS content, the prospects for synthesizing MPB compositions with very high Curie temperatures in this system are more promising.


\section{\label{sec:level1} Conclusions}

Using experimental data from the literature  and our own first-principles calculations, we have shown that the position of the MPB and $T_c$ at the MPB in PbTiO$_3$ based solid solutions can be  predicted from the average ionic radius and the B-cation displacement of the non-PbTiO$_3$ end member. Inclusion of first-principles data leads to significant improvements on the previously observed  empirical correlations, allowing quantitatively accurate predictions of MPB location and $T_c^{\rm MPB}$.  
 In agreement with previous work, large average B-cation radius (or lower tolerance factor) is found to favor high PT content at the MPB.  Large values of B-cation displacements give rise to lower PT content at the MPB.   Our model predicts that tetragonal phase will always be favored in BZT-PT solid solution and suggests that PZW-PT  solid solution will exhibit a phase digram different from other Pb(B$^{2+}$,B$^{6+}$)O$_3$-PT solutions. Based on the relationship between tolerance factor and $T_c^{\rm MPB}$, we suggest that BiZn$_{1/2}$Zr$_{1/2}$O$_3$-PT and BiCd$_{1/2}$Ti$_{1/2}$O$_3$-PT solid solutions  are the most promising candidates for the synthesis of high $T_c$ piezoelectrics.

This work was supported by the Office of Naval Research, under grant numbers N-000014-00-1-0372 and N00014-01-1-0860 and through the Center for Piezoelectrics by Design. We also acknowledge the support of the National Science Foundation, through the MRSEC program, grant No. DMR00-79909.   Computational support was provided by the Center for Piezoelectrics by Design, the  DoD HPCMO, DURIP and by the NSF CRIF program, Grant CHE-0131132.

\clearpage     
\bibliography{apssamp}


\appendix


\clearpage
\begin{table}[!t]
\caption{Results of our DFT calculations for ferroelectric compositions of PbTiO$_3$ solid solutions.  Ground-state cation displacements from center of oxygen cage in \AA.  Rhombohedral, monoclinic and tetragonal phases are denoted by $R$, $M$ and $T$ respectively.}

\begin{tabular}{lcccccc}

&\multicolumn{1}{c}{Pb,Bi}
&\multicolumn{1}{c}{B$'$}
&\multicolumn{1}{c}{B$''$}
&\multicolumn{1}{c}{\rm Ti}
&\multicolumn{1}{c}{}
&\multicolumn{1}{c}{Phase}\\ 

&\multicolumn{1}{c}{disp}
&\multicolumn{1}{c}{disp}
&\multicolumn{1}{c}{disp}
&\multicolumn{1}{c}{disp}
&\multicolumn{1}{c}{}
&\multicolumn{1}{c}{}\\ 

\hline
PMN-0.25PT &	0.389&	0.080&	0.162&	0.220&		& $R$\\	
PMN-0.63PT &	0.387&	0.099&	0.174&	0.233&		& $T$\\	
PT	   &	0.440&	     &	     &	0.280&		& $T$\\	

 	   &	& & &	&		&\\
	
PZN-0.25PT &	0.461&	0.258&	0.171&	0.272&		& $T$\\  
PZN-0.63PT &	0.424&	0.270&	0.187&	0.256&		& $T$\\	  
PT	   &	0.440&	     &	     &  0.280&		& $T$\\	
 	   &	& & &	&		&\\

PSW-0.25PT &	0.344&	0.095&	0.070&	0.241&		& $R$\\	
PSW-0.63PT &	0.350&	0.087&	0.126&	0.208&		& $T$\\	
PT	   &	0.440&	     &	     &	0.280&		& $T$\\	
 	   &	& & &	&		&\\

PZ-33PT    &	0.462&	0.129&	0.129&	0.235&		& $R$\\	
PZ-50PT    &	0.463&	0.138&	0.138&	0.247&		& $M$\\	
PZ-66PT    &	0.495&	0.175&	0.175&	0.278&		& $T$\\	
PT	   &	0.440&	     &	     &	0.280&		& $T$\\	
 &	& & &	&		&\\
PSN-0.25PT &	0.402&	0.112&	0.177&	0.236&		& $R$\\	
PSN-0.50PT &	0.381&	0.112&	0.155&	0.243&		& $M$\\	
PSN-0.75PT &	0.411&	0.143&	0.166 &	0.240&		& $T$\\	
PT	   &	0.440&	     &	     &	0.280&		& $T$\\	
 &	& & &	&		&\\
PIN-0.25PT &	0.407&	0.090&	0.156&	0.292&		& $R$\\	
PIN-0.50PT &	0.368&	0.073&	0.162&	0.272&		& $T$\\	
PIN-0.75PT &	0.385&	0.083&	0.178&	0.261&		& $T$\\	
PT	   &	0.440&	     &	     &	0.280&		& $T$\\

\end{tabular}
\end{table}

\clearpage
\begin{table}[!t]
\caption{Position of the MPB in PbTiO$_3$ solid solutions. Data for BMW-PT and PSW-PT solutions taken from Ref. 18 and 20 respectively.  Data for all other systems taken from Ref. 6.  End-members with high ferroelectric activity on the B-site are marked by an asterisk. For Pb(Mn,W)O$_3$-PT solid solution,  data is given for PbMn$_{2/3}$,W$_{1/3}$O$_3$ and PbMn$_{1/2}$,W$_{1/2}$)O$_3$ possible end-member stoichiometries.  For PbFe$_{2/3}$,W$_{1/3}$O$_3$-PT solid solution the average position of the MPB region that extends from 0.2 to 0.4 PT content~\cite{Mitoseriu03p1918} is given.}

\begin{tabular}{lcccccc}
&\multicolumn{1}{c}{}
&\multicolumn{1}{c}{$t$}
&\multicolumn{1}{c}{}
&\multicolumn{1}{c}{$x^{\rm MPB}_{\rm PT}$}
\\ 
&\multicolumn{1}{c}{}
&\multicolumn{1}{c}{}
&\multicolumn{1}{c}{}
&\multicolumn{1}{c}{}
\\ 

\hline
PbFe$_{2/3}$W$_{1/3}$O$_3$      &	&1.007&	&0.30\\	
PbMn$_{2/3}$W$_{1/3}$O$_3$      &	&1.007&	&0.53\\	
PbFe$_{1/2}$Nb$_{1/2}$O$_3$*      &	&1.001&	&0.10\\	
PbMn$_{1/2}$Nb$_{1/2}$O$_3$      &	&1.001&	&0.25\\	
PbNi$_{1/3}$Nb$_{2/3}$O$_3$      &	&0.994&	&0.35\\	
PbMg$_{1/2}$W$_{1/2}$O$_3$      &	&0.992&	&0.45\\	
PbMg$_{1/3}$Ta$_{2/3}$O$_3$      &	&0.989&	&0.32\\	
PbMg$_{1/3}$Nb$_{2/3}$O$_3$      &	&0.989&	&0.33\\	
PbCo$_{1/2}$W$_{1/2}$O$_3$      &	&0.986&	&0.45\\	
PbZn$_{1/3}$Nb$_{2/3}$O$_3$*      &	&0.986&	&0.10\\	
Na$_{1/2}$Bi$_{1/2}$TiO$_3$*      &	&0.980&	&0.14\\	
PbMn$_{1/2}$W$_{1/2}$O$_3$       &	&0.964& &0.53\\	
PbSnO$_3$	   &	&0.978&	&0.45\\	
PbSc$_{1/2}$Nb$_{1/2}$O$_3$      &	&0.977&	&0.43\\	
PbSc$_{1/2}$Ta$_{1/2}$O$_3$      &	&0.977&	&0.45\\	
PbHfO$_3$     &    &0.969&	&0.52\\
PbZrO$_3$      &	&0.964&	&0.48\\	
PbSc$_{2/3}$W$_{1/3}$O$_3$      &	&0.974&	&0.45\\	
PbIn$_{1/2}$Nb$_{1/2}$O$_3$      &	&0.964&	&0.36\\	
PbIn$_{1/2}$Ta$_{1/2}$O$_3$      &	&0.964&	&0.38\\	
BiFeO$_3$*      &	&0.956&	&0.28\\	
BiMnO$_3$      &	&0.956&	&0.50\\	
BiNi$_{1/2}$Ti$_{1/2}$O$_3$*      &	&0.955&	&0.45\\	

\end{tabular}
\end{table}

\clearpage
\begin{table}[!t]

\begin{tabular}{lcccccc}
&\multicolumn{1}{c}{}
&\multicolumn{1}{c}{$t$}
&\multicolumn{1}{c}{}
&\multicolumn{1}{c}{$x^{\rm MPB}_{\rm PT}$}
\\ 
&\multicolumn{1}{c}{}
&\multicolumn{1}{c}{}
&\multicolumn{1}{c}{}
&\multicolumn{1}{c}{}
\\ 

\hline

PbYb$_{1/2}$Nb$_{1/2}$O$_3$      &	&0.949&	&0.50\\	
BiMg$_{1/2}$Ti$_{1/2}$O$_3$*     &	&0.948&	&0.34\\	
BiMg$_{3/4}$W$_{1/4}$O$_3$      &	&0.934&	&0.50\\	
BiMg$_{1/2}$Zr$_{1/2}$O$_3$      &	&0.922&	&0.60\\	
BiScO$_3$      &	&0.911&	&0.65\\

\end{tabular}
\end{table}

\clearpage
\begin{table}[!t]
\caption{Ionic data and predictions for PT content at MPB using Eqns. 11-12 for Pb- and Bi-based end member perovskites. Shannon-Prewitt B-cation ionic radii and displacements obtained by DFT calculations are in \AA.  Fe displacement value is taken from Ref. 36. B-cation displacement data marked with an asterisk are estimated based on crystal chemical arguments. For Pb(Mn,W)O$_3$,  data is given for PbMn$_{1/2}$W$_{1/2}$O$_3$ and  PbMn$_{2/3}$W$_{1/3}$O$_3$  stoichiometries. }

\begin{tabular}{lcccccc}

&\multicolumn{1}{c}{B$'$,B$''$}
&\multicolumn{1}{c}{$t$}
&\multicolumn{1}{c}{B}
&\multicolumn{1}{c}{\rm MPB}
&\multicolumn{1}{c}{\rm MPB}
&\multicolumn{1}{c}{}\\ 

&\multicolumn{1}{c}{size}
&\multicolumn{1}{c}{}
&\multicolumn{1}{c}{disp}
&\multicolumn{1}{c}{predict}
&\multicolumn{1}{c}{exp}
&\multicolumn{1}{c}{}\\ 

\hline
PMW 	   &	0.72,0.60&	0.992&	0.08,0.10&      0.46& 0.45&\\	
PCoW       &    0.745,0.60&	0.986&	0.08*,0.10&	0.47& 0.45&\\	
PMnW       &	0.83,0.60&	0.966&	0.08*,0.10&	0.51& 0.53&\\	

 	   &	    &            &           &	         &	&\\

PMnN 	   &	0.645,0.64&	1.001&	0.12*,0.17&      0.28& 0.25&\\	
PFN        &	0.645,0.64&	1.001&	0.17,0.17&	0.16& 0.10&\\  
PSN        &	0.745,0.64&	0.977&	0.11,0.17&	0.37& 0.43& \\	  
PIN	   &	0.80,0.64 &	0.964&  0.07,0.17&	0.45& 0.36& \\	
PYN	   &	0.87,0.64 &	0.949&  0.08*,0.17&	0.48& 0.50&\\	
 	   &	& & &	&		&\\

PNN 	   &	0.69,0.64&	0.994&	0.08*,0.17&      0.32&  0.35&\\	
PMN        &	0.72,0.64&	0.989&	0.08,0.17&	0.33&  0.33&\\	  
PZN	   &	0.74,0.64&	0.986&  0.25,0.17&	0.09&  0.10&\\	
 	   &	& & &	&		&\\

PSW 	   &	0.745,0.60&	0.975&	0.11,0.10&	0.46& 0.45&\\	
PFW        &	0.645,0.60&	1.002&	0.17,0.10&	0.25&  0.30&\\	
PMnW       &	0.645,0.60&	1.007&	0.12*,0.10&	0.37&  0.53&\\	
 	   &	    &            &           &	         &	&\\
PZ         &	0.72,72&	0.964&	0.13,0.13&	0.43&  0.48&\\	
 	   &	    &            &           &	         &	&\\

BF 	   &	0.645,0.645&	0.956&	0.17,0.17&	0.29&  0.28&\\	
BMn        &	0.645,0.645&	0.956&	0.12*,0.12*&	0.44&  0.50&\\	
BS         &	0.745,0.745&	0.911&	0.11,0.11&	0.59&  0.65&\\	
 	   &	    &            &           &	         &	&\\

\end{tabular}
\end{table}
\clearpage
\begin{table}[!t]

\begin{tabular}{lcccccc}

&\multicolumn{1}{c}{B$'$,B$''$}
&\multicolumn{1}{c}{$t$}
&\multicolumn{1}{c}{B}
&\multicolumn{1}{c}{\rm MPB}
&\multicolumn{1}{c}{\rm MPB}
&\multicolumn{1}{c}{}\\ 

&\multicolumn{1}{c}{size}
&\multicolumn{1}{c}{}
&\multicolumn{1}{c}{disp}
&\multicolumn{1}{c}{predict}
&\multicolumn{1}{c}{exp}
&\multicolumn{1}{c}{}\\ 

\hline

BNiT 	   &	0.69,0.60&	0.955&	0.08*,0.25&      0.31&  0.45&\\	
BMgT       &	0.72,0.60&	0.948&	0.08,0.25&	0.35&  0.34&\\  
BMgZ       &	0.72,0.72&	0.922&	0.08,0.13&	0.57&  0.60&\\	  
 	   &	    &            &           &	         &	&\\
BMgW       &	0.72,0.60&	0.934&	0.08,0.10&	0.57&   0.48&\\	

 	   &	& & &	&		&\\

\end{tabular}
\end{table}

\clearpage
\begin{table}[!t]
\caption{$T_c$ at the MPB  in PbTiO$_3$ solid solutions. Data for for Bi(Mg,W)O$_3$-PT and Pb(Sc,W)O$_3$-PT solid solutions taken from Ref. 18 and 24 respectively. Solid solutions with lower ferroelectric activity on the B-site are marked by an asterisk. }

\begin{tabular}{lcccccc}
&\multicolumn{1}{c}{}
&\multicolumn{1}{c}{t}
&\multicolumn{1}{c}{}
&\multicolumn{1}{c}{$T_c^{\rm MPB}$}
\\ 
&\multicolumn{1}{c}{}
&\multicolumn{1}{c}{}
&\multicolumn{1}{c}{}
&\multicolumn{1}{c}{}
\\ 

\hline

Pb(Co,W)O$_3$*      &	&0.987     &	&30\\	
Pb(Fe,W)O$_3$      &	&1.001     &	&52\\	
Pb(Mg,W)O$_3$*      &	&0.993     &    &60\\	
Pb(Mg,TaO)$_3$      &	&0.989     &	&80\\	
Pb(Sc,W)O$_3$*      &	&0.975     &	&97\\	
Pb(Ni,Nb)O$_3$      &	&0.994     &	&100\\	
Pb(Fe,Nb)O$_3$      &	&1.001     &	&130\\	
Pb(Mg,Nb)O$_3$      &	&0.989     &	&140\\	
Pb(Mn,Nb)O$_3$      &	&0.973     &	&160\\	
Pb(Zn,Nb)O$_3$      &	&0.986     &	&187\\	
Pb(Sc,Ta)O$_3$      &	&0.977     &	&190\\	
PbSnO$_3$      &	&0.978     &	&205\\	
Pb(Co,Nb)O$_3$      &	&0.985     &	&220\\	
Bi(Mg,W)O$_3$*     &	&0.933     &	&225\\	
Pb(Sc,Nb)O$_3$     &	&0.977     &	&260\\	
Bi(Mg,Zr)O$_3$*      &	&0.921     &	&300\\	
Pb(In,Nb)O$_3$      &	&0.965     &	&320\\	
PbHfO$_3$      &	&0.969     &	&340\\	
(Na,Bi)TiO$_3$      &	&0.977     &	&350\\	
Pb(Yb,Nb)O$_3$      &	&0.951     &	&360\\	
PbZrO$_3$      &	&0.964     &	&385\\	
Bi(Ni,Ti)O$_3$      &	&0.955     &	&425\\	
Bi(Mg,Ti)O$_3$      &	&0.947     &	&425\\	
BiScO$_3$*      &	&0.911     &	&450\\

\end{tabular}
\end{table}

\clearpage
\begin{table}[!t]
\caption{Predicted $x_{\rm PT}^{\rm MPB}$ and $T_c^{\rm MPB}$ values for PbB$'$B$''$O$_3$-PT and BiB$'$B$''$O$_3$-PT solid solutions discussed in Section IV.  Predicted $x_{\rm PT}^{\rm MPB}$ are obtained using Eqn. 11-12, $T_c^{\rm MPB}$ (in $^{\circ}$C) are obtained by uisng Eqn. 13-14.  Ionic size and displacement data in \AA.  }

\begin{tabular}{lcccccc}

&\multicolumn{1}{c}{B$'$,B$''$}
&\multicolumn{1}{c}{B$'$,B$''$}
&\multicolumn{1}{c}{$t$}
&\multicolumn{1}{c}{Avg B}
&\multicolumn{1}{c}{$x^{\rm MPB}_{\rm PT}$}
&\multicolumn{1}{c}{$T_c^{\rm MPB}$}\\ 

&\multicolumn{1}{c}{size}
&\multicolumn{1}{c}{disp}
&\multicolumn{1}{c}{}
&\multicolumn{1}{c}{disp}
&\multicolumn{1}{c}{predict}
&\multicolumn{1}{c}{predict}\\ 

\hline
PbZn$_{1/2}$W$_{1/2}$O$_3$        &	0.74,0.60&	0.25,0.10& 0.987&	0.20 &	0.13& 180\\	
PbCd$_{1/2}$W$_{1/2}$O$_3$        &	0.95,0.60&	0.05,0.10& 0.940&	0.08 &	0.55& 487 \\	
 	   &	    &             &           &	         &	&\\
BiInO$_3$                         &	0.80,0.80&	0.07,0.07& 0.887&	 0.05&	0.67& 507 \\	
BiYbO$_3$                         &	0.86,0.86&	0.08,0.08& 0.863&	 0.05&	0.70& 615 \\	
 	   &	    &            &           &	         &	&\\
BiCd$_{1/2}$Ti$_{1/2}$O$_3$       &	0.95,0.60&	0.08,0.25& 0.897&	0.15&	0.54&   770\\	
BiZn$_{1/2}$Zr$_{1/2}$O$_3$       &	0.74,0.72&	0.25,0.13& 0.916&	0.19&	0.42&   644\\	
BiZn$_{2/3}$Nb$_{1/3}$O$_3$       &	0.74,0.64&	0.25,0.17& 0.926&	0.22&	0.26&   579\\	
BiZn$_{3/4}$W$_{1/4}$O$_3$       &	0.74,0.60&	0.25,0.10& 0.927&	0.21&	0.30&   572\\

\end{tabular}
\end{table}

\clearpage
\begin{figure}
\includegraphics[width=4.0in]{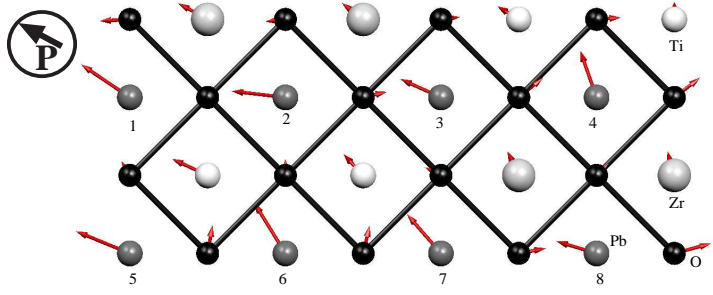}
\caption{{(Color online). Projection of the DFT relaxed structure for $4\times 2 \times 1$ 50/50 PZT supercell on the $x-y$ plane. Pb atoms tend to distort away from the large  Zr and toward smaller Ti cations.}}
\end{figure}

\begin{figure}
\centering
\includegraphics[width=3.0in]{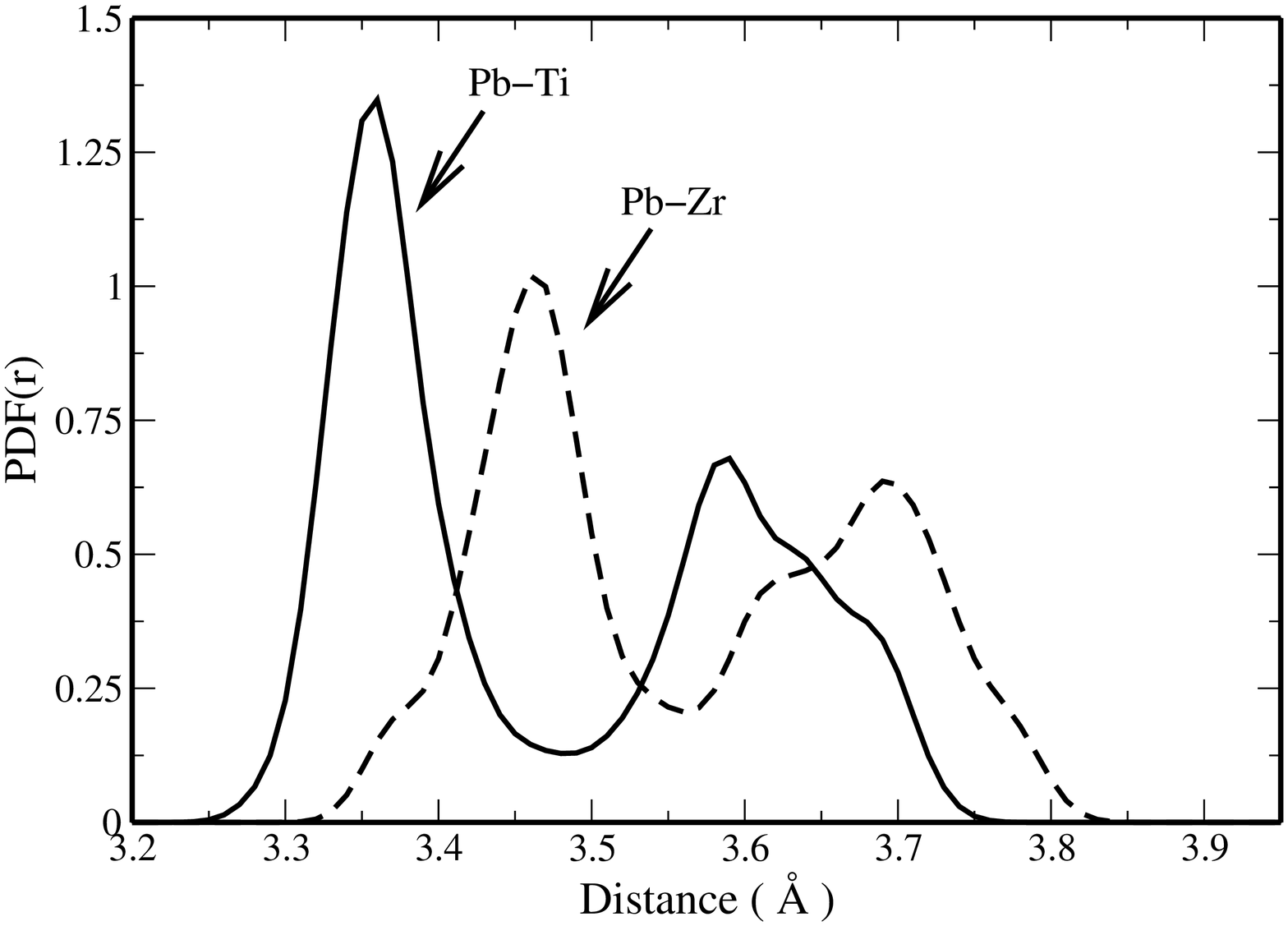}
\includegraphics[width=3.0in]{PbScIn.eps}
\caption{{ Pb-B-cation partial PDFs.  (a) Pb-Ti (solid) and Pb-Zr (dashed) partial PDFs obtained from the relaxed structures of the 50/50 PZT supercells at experimental lattice constants. (b) Pb-Sc (solid) and Pb-In (dashed) partial PDFs obtained from the relaxed structures of PSN and PIN at experimental lattice constants.}}
\end{figure}

\begin{figure}
\includegraphics[width=4.5in]{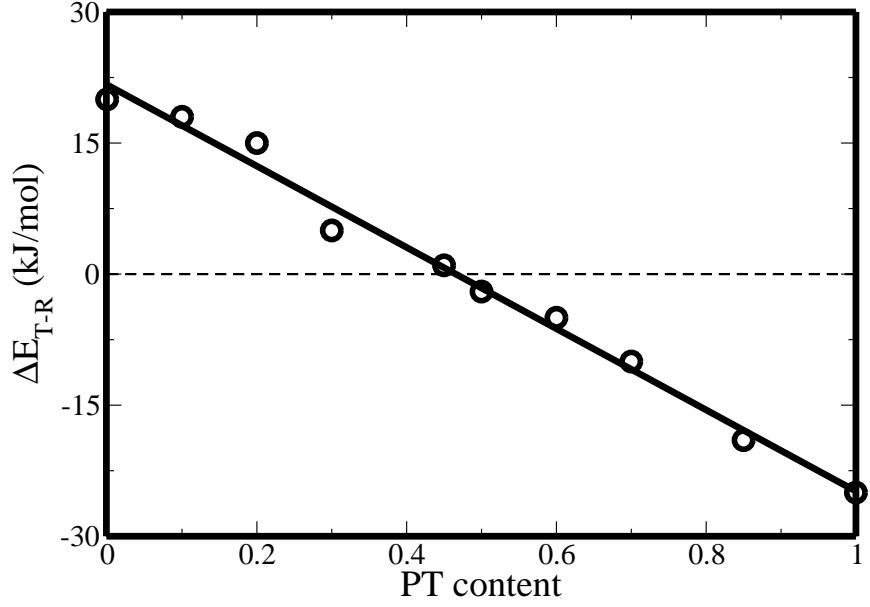}
\caption{{ Experimental $\Delta E_{T-R}$ values (circles) for the PZT solid solutions as function of Ti composition~\cite{Rossetti02p8131}. The dependence of $\Delta E_{T-R}$ on PT is well fit by a linear function for the whole  compositional range.}}
\end{figure}

\begin{figure}
\includegraphics[width=4.5in]{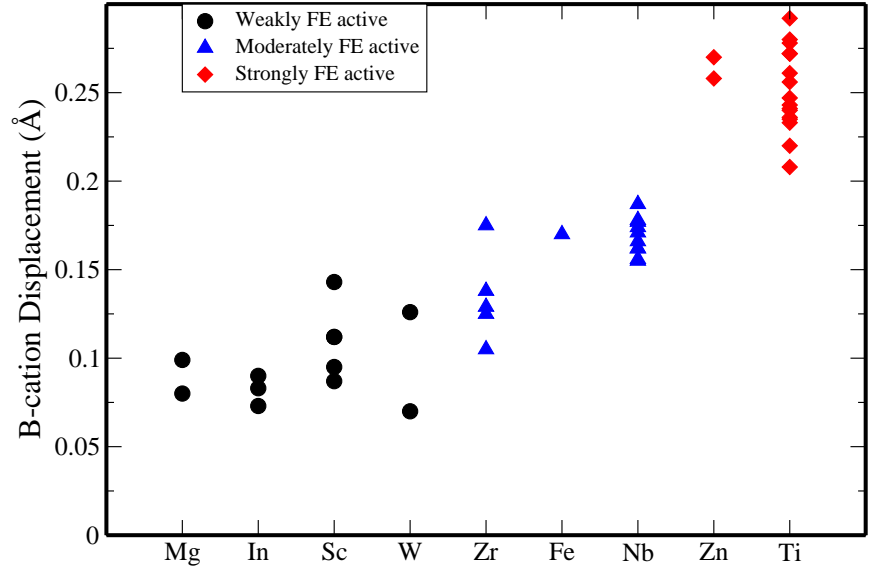}
\caption{{(Color online). Distortion magnitudes (in \AA) for B-cations in PT solid solutions. Distortion magnitudes are especially large for Zn and Ti.  $D_{\rm B}$ values of 0.08~\AA, 0.07~\AA, 0.11~\AA, 0.10~\AA, 0.13~\AA, 0.17~\AA, 0.17~\AA, 0.25~\AA$ $ and 0.25~\AA $ $ are assigned for Mg, In, Sc, W, Zr, Fe, Nb, Zn and Ti respectively.}}
\end{figure}

\begin{figure}
\includegraphics[width=4.5in]{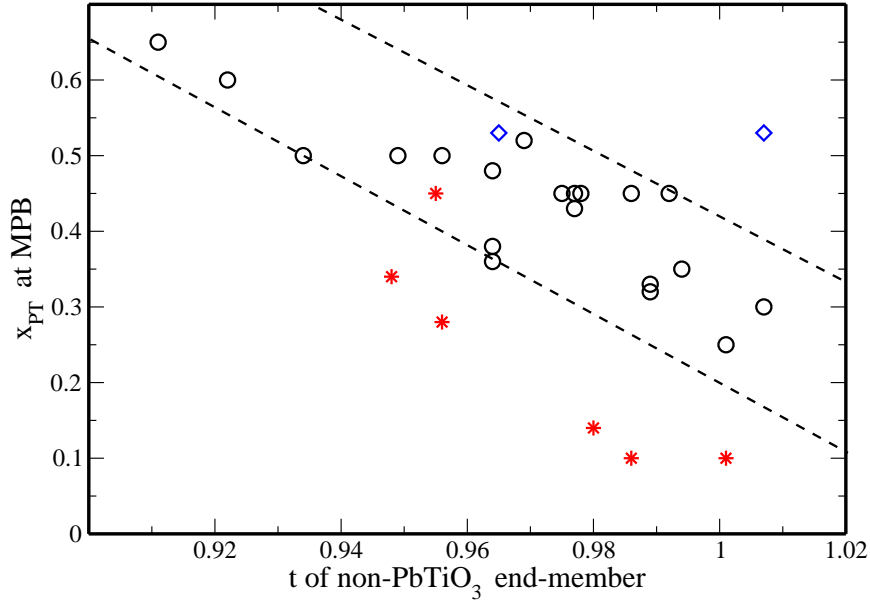}
\caption{{(Color online). Plot of the MPB position (in mole fraction PT)  versus the end member tolerance factor.   Circles indicate non-PT end-member perovskites with no Ti-cations and less than full occupation of the B-site by ferroelectrically active Nb or Fe cations. These display a lower ferroelectric activity on the end-member B-site. Stars indicate systems where ferroelectric activity of the non-PT end-member B-site is higher due to presence of Ti ions and/or occupation of  more than one half of the B-sites by ferroelectrically active Fe and Nb ions.  Diamonds indicate data for Pb(Mn,W)O$_3$-PT solid solution with 1:1 (low tolerance factor) and 2:1 Mn/W (high tolerance factor) stoichiometries.}}
\end{figure}

\begin{figure}
\includegraphics[width=4.0in]{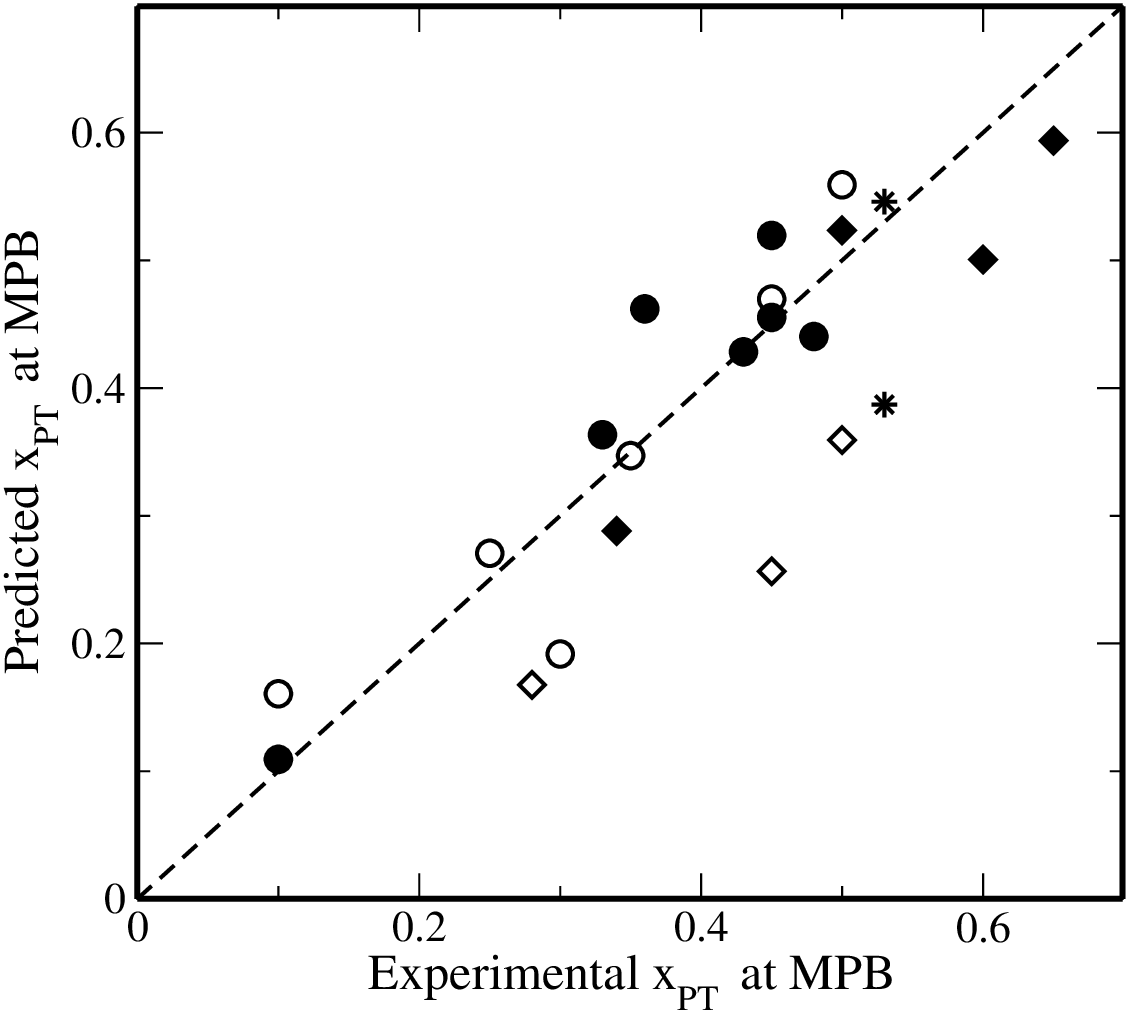}
\caption{{ Correlation between the mole fractions of PT at MPB predicted by Eqn. 10 using data in Table III and  mole fractions of PT at MPB observed experimentally. Solid solutions for which we have DFT B-cation displacement data are marked by filled cirles and diamonds for Pb-based and Bi-based systems, respectively.  MPB positions predicted for Pb- and Bi-based systems where B-cation displacement data is estimated are marked by open circles and  diamonds respectively. Predicted positions for MPB for Pb(Mn,W)O$_3$-PT solid solutions  using 1:1 (predicted $x^{\rm MPB}_{\rm PT}$=0.55)    and 2:1 (predicted $x^{\rm MPB}_{\rm PT}$=0.39) Mn:W stoichiometries are represented by stars. Taking the effect of B-cation off-centering into account makes quantitative prediction of MPB location possible. }}
\end{figure}

\begin{figure}
\includegraphics[width=4.0in]{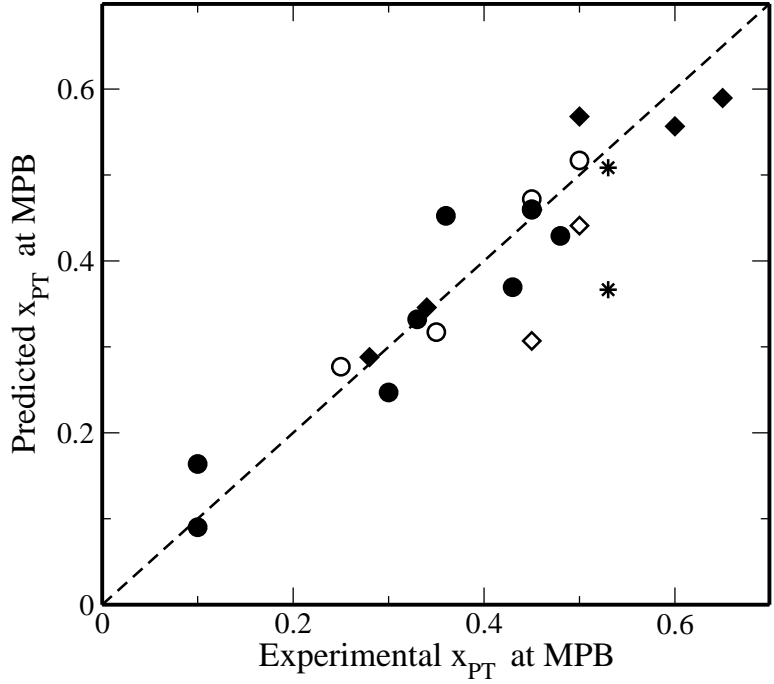}
\caption{{ Correlation between the mole fractions of PT at MPB predicted by Eqn. 11-12 using data in Table III and  mole fractions of PT at MPB observed experimentally.  Solid solutions for which we have DFT B-cation displacement data are marked by filled cirles and diamonds for Pb-based and Bi-based systems respectively.  MPB positions predicted for Pb- and Bi-based systems where B-cation displacement data is estimated are marked by open circles and  diamonds respectively. Predicted positions for MPB for Pb(Mn,W)O$_3$-PT solid solutions  using 1:1 (predicted $x^{\rm MPB}_{\rm PT}$=0.51)    and 2:1 (predicted $x^{\rm MPB}_{\rm PT}$=0.37) Mn:W stoichiometries are represented by stars. Using separate fits for Pb- and Bi-based end-member perovskites improves the correlation.}}
\end{figure}

\begin{figure}
\includegraphics[width=4.0in]{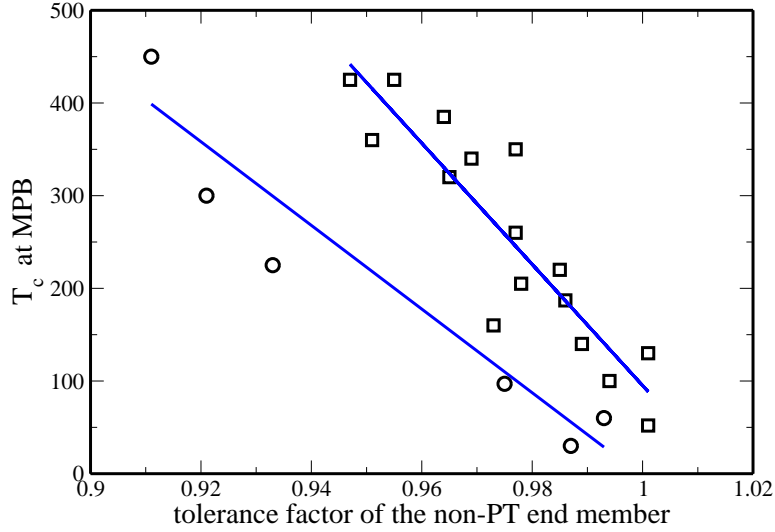}
\caption{{(Color online). $T_c^{\rm MPB}$ versus tolerance factor of the non-PT end-member using data of Table V. End-members with lower FE activity on the B-site are shown by circles, end-members with higher FE activity on the B-site  are shown by squares.  Low concentration of ferroelectrically active cations on the end-member B-site shifts the $T_c^{\rm MPB}$ to lower values.  Least-squares fit lines for the two cases are shown. }}
\end{figure}

\begin{figure}
\includegraphics[width=4.0in]{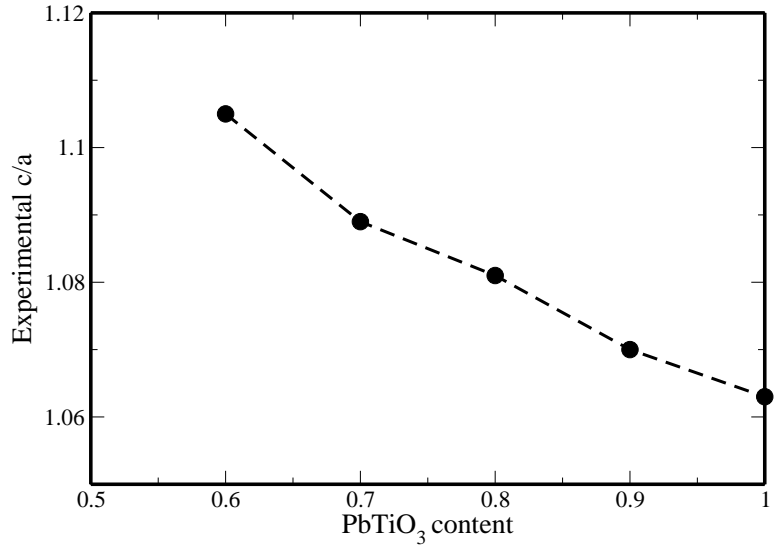}
\caption{{ Experimental data for $c/a$ ratio in BiZn$_{1/2}$Ti$_{1/2}$O$_3$-PT solid solution.  The material is tetragonal for all compositions studied, with an anomalous strong enhancement in tetragonality.  Compositions with less then 0.6 PT content could not be made as a single phase. }}
\end{figure}

\end{document}